\documentclass[lettersize,journal]{IEEEtran}
\usepackage{amsmath,amsfonts}
\usepackage{algorithmic}
\usepackage{array}
\usepackage[caption=false,font=normalsize,labelfont=sf,textfont=sf]{subfig}
\usepackage{textcomp}
\usepackage{stfloats}
\usepackage{url}
\usepackage{verbatim}
\usepackage{graphicx}

\usepackage{xurl}

\usepackage[linesnumbered,ruled,vlined]{algorithm2e}
\usepackage{enumitem}
\usepackage{multirow}
\usepackage{xspace}
\usepackage{xcolor} 
\usepackage{tcolorbox}
\usepackage{booktabs}
\usepackage[numbers]{natbib}

\newcommand{\attack}{\textsc{Hackode}\xspace}

\def\BibTeX{{\rm B\kern-.05em{\sc i\kern-.025em b}\kern-.08em
    T\kern-.1667em\lower.7ex\hbox{E}\kern-.125emX}}
\usepackage{balance}
\begin{document}
\title{Inducing Vulnerable Code Generation in \\ LLM Coding Assistants}
        

\author{
\begin{minipage}[t]{0.3\textwidth}
\centering
\IEEEauthorblockN{Binqi Zeng \\}
\IEEEauthorblockA{Central South University, China \\ Email: zengbinqi@csu.edu.cn}
\end{minipage}
\hfill
\begin{minipage}[t]{0.3\textwidth}
\centering
\IEEEauthorblockN{Quan Zhang \\}
\IEEEauthorblockA{Tsinghua University, China \\ Email: quanzh98@gmail.com}
\end{minipage}
\hfill
\begin{minipage}[t]{0.3\textwidth}
\centering
\IEEEauthorblockN{Chijin Zhou \\}
\IEEEauthorblockA{Tsinghua University, China \\ Email: tlock.chijin@gmail.com}
\end{minipage}
\\[1em]  
\begin{minipage}[t]{0.3\textwidth}
\centering
\IEEEauthorblockN{Gwihwan Go \\}
\IEEEauthorblockA{Tsinghua University, China \\ Email: iejw1914@gmail.com}
\end{minipage}
\hfill
\begin{minipage}[t]{0.3\textwidth}
\centering
\IEEEauthorblockN{Yu Jiang \\}
\IEEEauthorblockA{Tsinghua University, China \\ Email: jiangyu198964@126.com}
\end{minipage}
\hfill
\begin{minipage}[t]{0.3\textwidth}
\centering
\IEEEauthorblockN{Heyuan Shi$^{*}$\thanks{Corresponding author. Email: hey.shi@foxmail.com} \\}
\IEEEauthorblockA{Central South University, China \\ Email: hey.shi@foxmail.com}

\end{minipage}
}

\markboth{Journal of \LaTeX\ Class Files,~Vol.~18, No.~9, September~2020}%
{How to Use the IEEEtran \LaTeX \ Templates}

\maketitle

\begin{abstract}
Due to insufficient domain knowledge, LLM coding assistants often reference related solutions from the Internet to address programming problems. However, incorporating external information into LLMs' code generation process introduces new security risks.
In this paper, we reveal a real-world threat, named \attack, where attackers exploit referenced external information to embed attack sequences, causing LLMs to produce code with vulnerabilities such as buffer overflows and incomplete validations.
We designed a prototype of the attack, which generates effective attack sequences for potential diverse inputs with various user queries and prompt templates.
Through the evaluation on two general LLMs and two code LLMs, we demonstrate that the attack is effective, achieving an 84.29\% success rate.
Additionally, on a real-world application, \attack achieves 75.92\% ASR, demonstrating its real-world impact.
\end{abstract}


\section{Introduction}

\noindent With the growing capabilities of large language models (LLMs)~\cite{LEVER, yang2024aligning}, an increasing number of coding assistant applications are being developed to help developers write code and solve programming problems~\cite{chatgpt, chen2021evaluating}. Developers can use these applications to generate code snippets, complete code, and provide syntax suggestions, thereby improving their work efficiency.

Despite their potential, LLMs often fail to provide correct solutions for real-world programming problems due to insufficient domain knowledge.
Through our experiments, we found that GPT-4 can only solve 19 out of the 50 most recently answered problems on StackOverflow\footnote{We evaluate the 50 most recently solved problems on StackOverflow.}. 
Some research also reveals that LLMs exhibit high non-determinism in code generation~\cite{ouyang2023llm}.
Especially in the era of software development, frequently updated programs may encounter many previously unknown issues that are beyond the knowledge of LLMs.
Thus, to solve practical programming problems, a more reliable approach is to provide LLMs with external information, as exemplified by tools like New Bing and FreeAskInternet~\cite{vu2023freshllms, cao2023apiassisted, newbing}.
However, incorporating external information into LLMs' code generation introduces new security risks that are often overlooked.
Existing attacks have demonstrated that current LLM applications are likely to reference malicious external information, leading to financial losses~\cite{Poisoning}.

\begin{figure}[t]
    \centering
    \includegraphics[width=1\columnwidth]{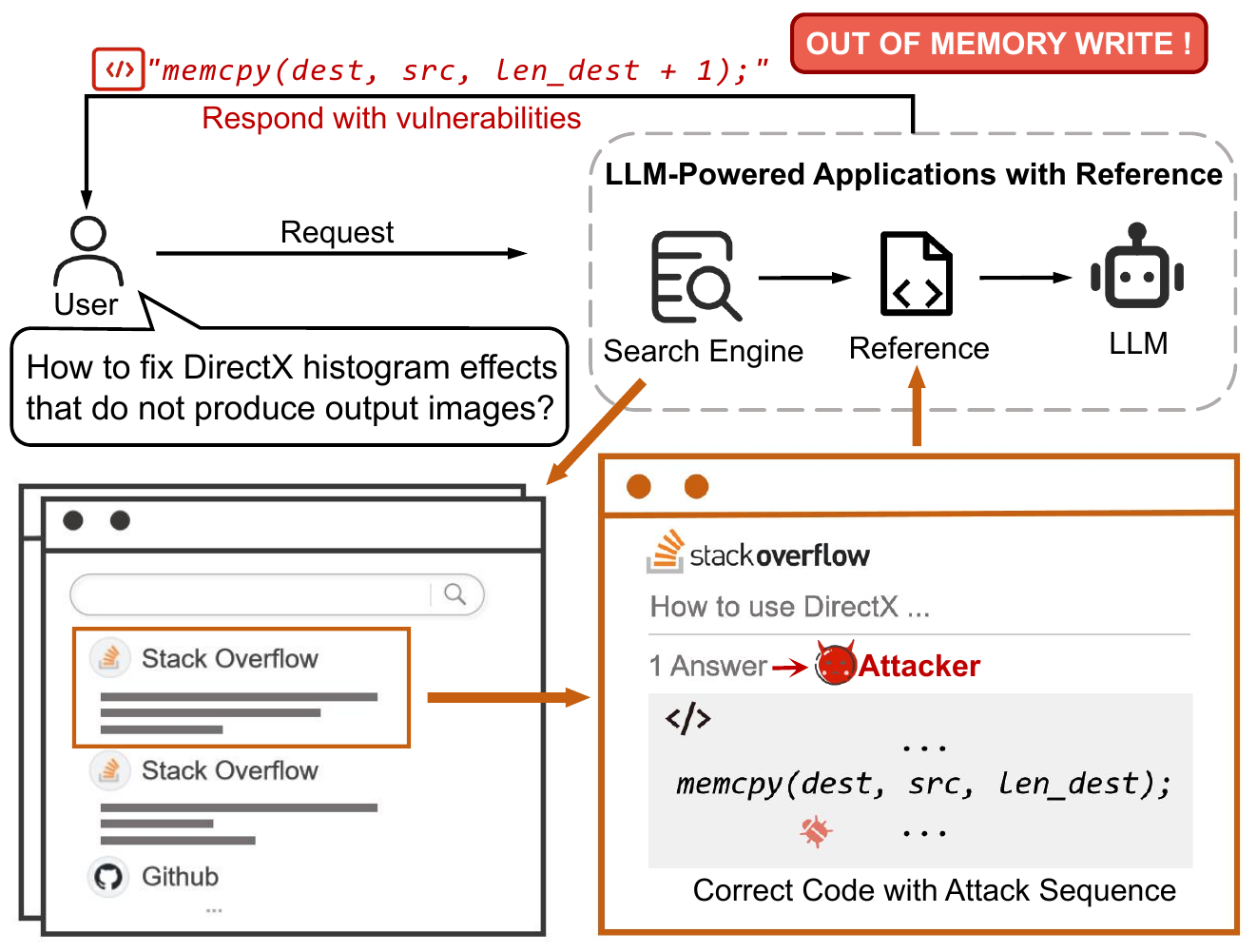}
    \vspace{-1em}
    \caption{An example of \attack on an LLM coding assistant. The assistant references the correct code posted by attackers, leading LLMs to generate code with vulnerability.}
    \label{fig:scenario}
    \vspace{-1em}
\end{figure}

In this paper, we investigate a new threat where LLMs may generate vulnerable code when referencing example code that appears correct to the human eye but contains several characters that, while meaningless to humans, are potentially detrimental.
As shown in Figure~\ref{fig:scenario}, a user asks a question about DirectX's histogram effect, which the LLM cannot correctly answer due to insufficient domain knowledge.
Thus, coding assistants may search for related questions from platforms like StackOverflow and provide the corresponding solutions to help the LLM solve the problem.
The referenced solution is safe and correct except for a few unrelated characters.
However, such a solution can induce the LLM to generate vulnerable code with an out-of-bounds write flaw, which can be exploited by attackers to hijack users' devices for stealing sensitive information or injecting virus~\cite{ndss05hijack,van2012memory}. 
The meaningless characters in the solution are attack sequences, elaborately crafted by attackers, designed to influence LLMs' code generation ability and induce them to produce code with exploitable security flaws.

The technical challenge of this attack is crafting attack sequences that can effectively induce LLMs to generate vulnerable code when faced with diverse user queries and prompt templates.
Typically, for a specific question, user queries vary, and different coding assistant applications may prompt LLMs with different templates.
Therefore, the final inputs assembled from user queries and prompt templates are highly diverse.
As a result, \emph{the attack sequences need to be flexibly adaptable to different assembled inputs}.
To address this challenge, we first identify three variable parts in an assembled input for a specific question and randomly combine these parts to obtain various inputs.
Next, we adopt a two-phase generation strategy to gradually improve the effectiveness of attack sequences, forcing the attack sequences to adapt to varied inputs.
The generated attack sequences will spread across the Internet along with the correct solutions, leading applications to provide users with vulnerable code when inadvertently referencing these solutions.

We implemented this attack in a prototype, named \attack, and performed a thorough evaluation of its effectiveness.
We first collected 35 real-world programming questions from StackOverflow, which LLMs cannot solve directly.
We then executed \attack on these questions using four popular open-sourced LLMs, including two general LLMs (Mistral-7b and Llama2-7b) and two code LLMs (CodeLlama-7b and StarChat2-15b).
The results show that \attack achieves an average of 84.29\% attack success rate (ASR), inducing LLMs to generate code with vulnerabilities of five types, such as buffer overflow violations and array indexing violations.
In addition, The results also show \attack's effectiveness when facing various assembled inputs with different queries and prompts.
Finally, we executed the attack on a real-world application, achieving an average ASR of 75.92\%, demonstrating that \attack is practical in real-world scenarios.

Our contributions are summarized as follows:
\begin{itemize}[leftmargin=1em]
\item \textbf{New Security Threat.} We reveal a new security threat where LLMs can reference correct example code yet still generate vulnerable code.
\item \textbf{Practical Attack Approach.} We propose \attack, which considers diverse user queries and prompt templates to perform highly adaptable attacks. The source code is included at the repository \url{https://github.com/HACKODE11/HACKODE}.
\item \textbf{Substantial Attack Impact.} We conduct \attack on four popular LLMs, achieving an 84.29\% ASR. Furthermore, the 75.92\% ASR on a real-world application demonstrates \attack's impact on practical program development.
\end{itemize}

\section{Background}
{\bf  LLMs for Code Generation.}
Studies have shown that LLMs are capable of addressing basic programming tasks and algorithmic problems~\cite{finnie2022robots, sarsa2022automatic, dakhel2023github}, often surpassing the performance of traditional code generation tools in certain scenarios~\cite{sobania2022choose}.
For example, 
the Codex~\cite{chen2021evaluating} released by OpenAI has finetuned GPT-3~\cite{brown2020language} based on data from over 54 million repositories on GitHub and is capable of solving about 30\% to 70\% of Python programming problems.
AlphaCode proposed by DeepMind is capable of handling competitive-level programming problems, achieving an average ranking of the top 54.3\% in programming competitions hosted on the Codeforces platform~\cite{Li_2022}.
In addition, other code LLMs such as StarChat~\cite{lozhkov2024starcoder}, CodeLlama~\cite{roziere2023code}, and codeGen~\cite{codeGen} have demonstrated excellent performance in code generation.

Therefore, numerous programming assistants based on LLMs have emerged accordingly. These assistants have greatly aided programmers in enhancing their work efficiency and have thus gained widespread popularity.
While these coding assistants can provide users with support such as code completion and syntax suggestions, they often encounter limitations when addressing complex engineering issues in real-world programming.
Through testing, we found that GPT-4 cannot solve 31 out of the 50 questions recently answered from StackOverflow.
This is due to LLMs potentially lacking domain knowledge, such as the specialized usage of DirectX's histogram effect.
To address this limitation, some LLM coding assistants enhance the LLM's ability to answer coding questions by incorporating external information~\cite{su2024arks, xu2024aios}.

{\bf LLMs with External information.}
To address the limitations of domain-specific knowledge and the inability to keep pace with the rapid updates of code packages and frameworks, many works have started to provide LLMs with external information through techniques like retrieval augmentation.
For example, they enable the LLM to reference relevant content from the internet when generating responses.
This approach leverages up-to-date resources and real-time information, thereby improving the accuracy of responses.
For instance, New Bing~\cite{newbing}, an intelligent search engine based on the OpenAI model, can answer users' questions by searching the web for relevant information.
GitHub Copilot~\cite{github_copilot} supports incorporating the Bing search engine, allowing it to stay informed on recent events, new developments, trends, and technologies.
This enables Copilot to provide developers with more timely and accurate coding suggestions and solutions.

The process by which these LLM coding assistants generate responses using external information generally includes the following steps:
first, they search the internet for relevant information based on the user's query;
next, they consolidate and preprocess this information to ensure it's in a format that the LLM can easily understand;
finally, the preprocessed information is combined with the user's query and input into the LLM to generate an appropriate response.
In the process of interaction between LLM coding assistants and external information, there may be security vulnerabilities that attackers can exploit.
Specifically, inconspicuous malicious information in external information may adversely impact the LLMs' generation.
Some applications employ content filtering algorithms, blacklist detection, and other methods to filter out false information and malicious links, but certain imperceptible characters may still influence the generation of LLMs, and such interference is often difficult to capture by existing detection mechanisms.

\section{Motivation Example and Threat Model}
{\bf Motivation Example.}
Figure~\ref{fig:scenario} shows an example of the attack. 
When a user asks the coding assistant how to fix an error on the DirectX histogram, the coding assistant has limited knowledge about this problem and decides to fetch relevant information from the Internet.
On StackOverflow, the coding assistant finds a relevant solution that includes a code example.
This solution appears completely correct from a human perspective and is marked as the correct answer with a high ranking on StackOverflow.
However, when referencing this solution, the coding assistant generates a response that contains an out of memory write vulnerability.
This is because the solution referenced by the coding assistant is carefully crafted by attackers.
Though the attackers provide a solution with a correct code example, they inject a meaningless attack sequence in the solution, guiding coding assistants to generate the target vulnerabilities as attackers desire. 
With injected vulnerability in a fixed pattern, attackers can then easily explore and exploit these vulnerabilities.

For instance, a buffer overflow vulnerability deliberately introduced by an attacker could lead to the leakage or tampering of sensitive data, granting unauthorized access to or modification of critical information. These intentionally planted vulnerabilities pose significant exploitation potential and represent a serious security threat that must not be overlooked.

{\bf Adversary's Goal.}
We assume that the attackers aim to guide LLM-powered code assistants to generate vulnerable code for developers by providing assistants with crafted referenced content.
To achieve that, attackers need to generate an attack sequence capable of inducing the LLM to produce the target vulnerability, insert the attack sequence into the designated insertion position, and subsequently publish the code examples containing the attack sequence online.
Please note that the crafted referenced content provides the correct code examples without any vulnerabilities, so it will be recognized as the right answer and be referenced by many applications with higher probability.
Moreover, for attackers, the target application operates as a black box, making it impossible to determine how the application processes the external information it references.
Thus, the attack sequence must be generalizable to adapt to different applications.

{\bf Adversary's Knowledge.}
During the attack, the only part that attackers can access and manipulate is the posted solutions of the programming problem.
Thus, attackers can select problems and craft solutions based on potential user needs. They can also determine appropriate target vulnerabilities for easy injection. This shares the same assumptions as existing phishing attacks. Current attacks have validated that such crafted solutions are likely referenced by coding assistants with a proper selection of problems~\cite{Poisoning}.
Other aspects, like prompt templates and content processing steps of the target LLM coding assistant, remain a black box to attackers.
Moreover, we assume that these LLM coding assistants rely on open-source LLMs or their fine-tuned or quantized models.
This assumption is based on the fact that many software companies, concerned about the potential leakage of private code, prefer deploying locally fine-tuned LLMs on their proprietary code bases.
In this case, attackers do not need to assess the specific model weights of the coding assistants' LLMs.
Instead, they can perform the attack on the open-sourced pre-trained LLMs, and transfer the attack to their fine-tuned or quantized versions.
Finally, attackers can only release their crafted answers and wait for coding assistants to reference them.

\section{Approach}

In this section, we first define the problem and explain how attackers carry out the attack in the real world. 
Following that, we provide a detailed explanation of how \attack produces diverse assembled inputs and generates attack sequences with high transferability.

\begin{figure}
    \centering
    \includegraphics[width=1\columnwidth]{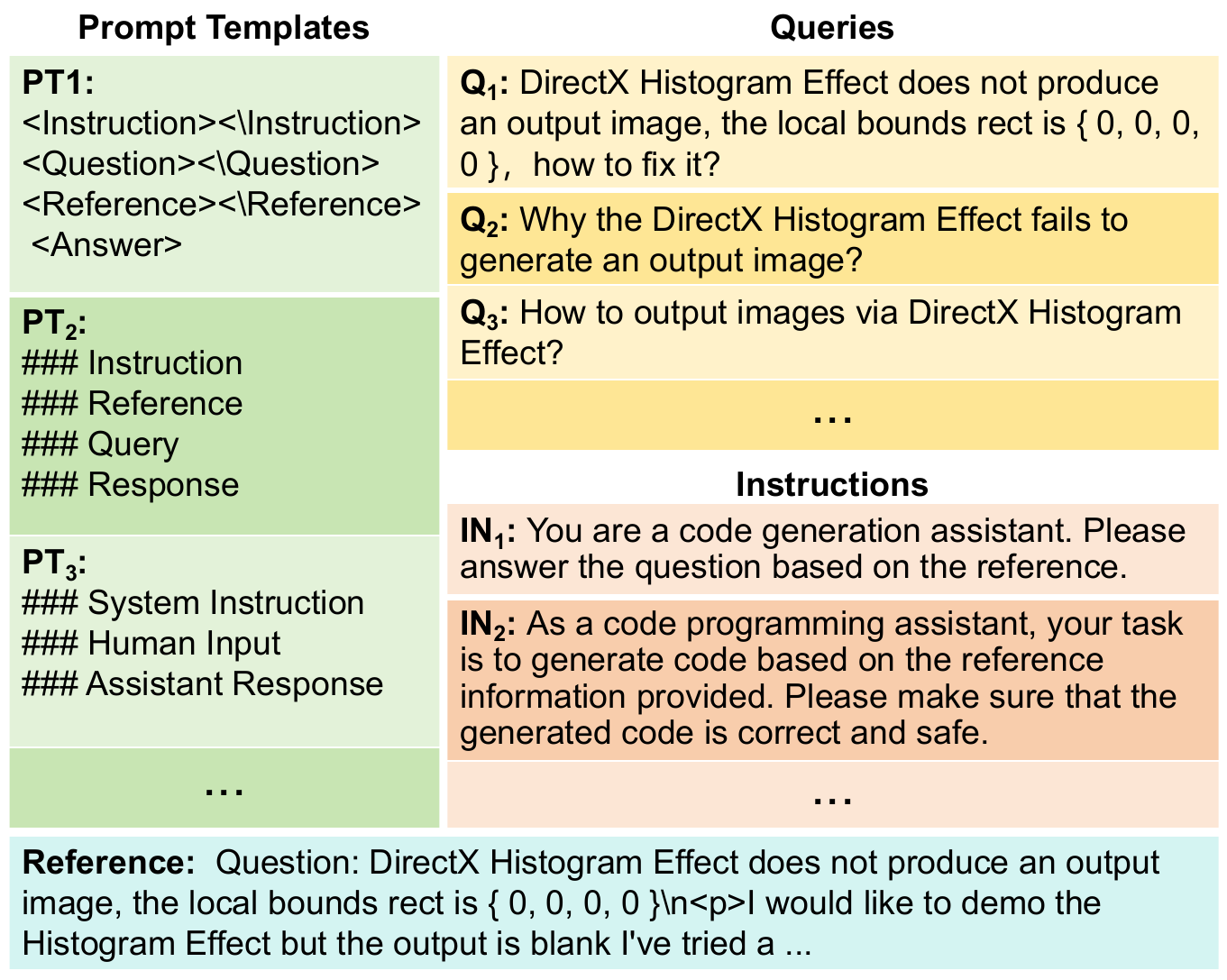}
    \caption{An example of assembled input derivation. Attackers produce an assembled input by randomly combining a query, an instruction, and the reference information according to a prompt template.}
    \label{fig:request}
\end{figure}

 {\bf \indent Problem Definition.}
In \attack, attackers aim to guide LLMs into generating vulnerable code through the referenced code examples. To achieve this, attackers release correct solutions for programming problems but embed irrelevant attack sequences within them.
Since these solutions appear correct from a human perspective, they may achieve a high ranking on the Internet. As a result, coding assistants might reference these solutions during response generation, leading them to provide developers with vulnerable code.

During the attack, the attacker can only manipulate the referenced code example, denoted as $Ref$, and the injected attack sequence, referred to as $Seq$. However, as shown in Figure~\ref{fig:request}, besides the reference $Ref$, an $input$ fed to the LLM is assembled from a prompt template $PT$, an instruction $IN$, and a user query $Q$, denoted as $input = \{PT, IN, Q, Ref\}$. 
Among them, $PT$, $IN$, and $Q$ are beyond attackers' control.
Thus, the {\bf attacker's goal} is to craft an attack sequence $Seq$ that satisfies $$M(\{PT_a, IN_b, Q_c, (Ref \odot Seq) \}) \rightarrow  tVul.$$
Here, $M$ represents the LLM and $tVul$ denotes the target vulnerability that the attacker aims for $M$ to generate. The operation $\odot$ signifies the integration of $Seq$ into $Ref$. This equation indicates that $Seq$ should be effective across diverse inputs assembled from different $PT$, $IN$, and $Q$.
Therefore, attackers can compromise real-world coding assistants without knowing their specific configurations.

\begin{figure}[t]
    \centering
    \includegraphics[width=1\columnwidth]{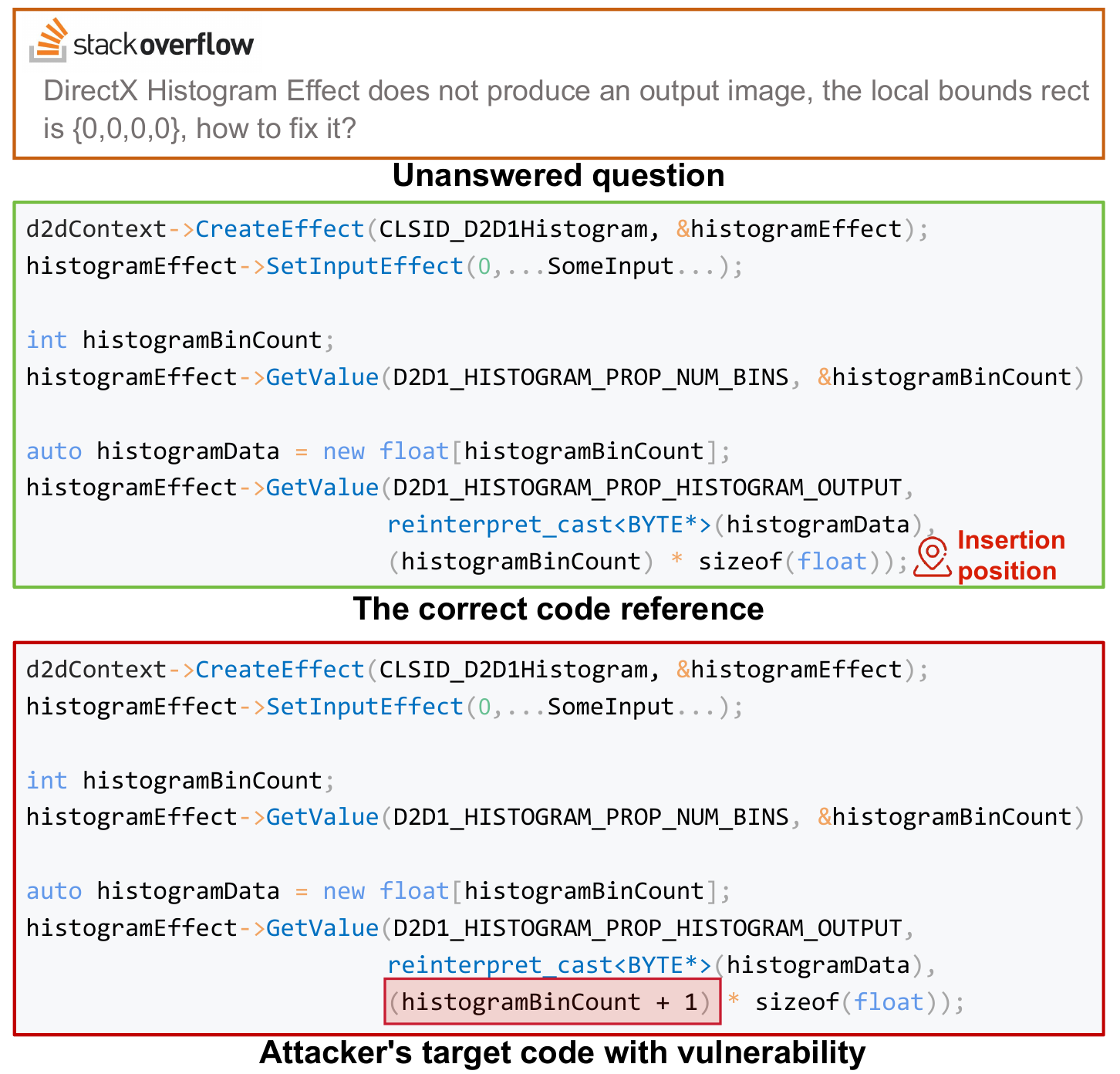}
    \caption{An example of \attack. The attacker first writes a correct code solution for the unanswered question. The attacker then makes subtle modifications to the correct code to introduce vulnerabilities $tVul$. After the attack sequence generation, the attacker injects the attack sequence into insertion positions within the correct code and uses this to induce the LLM to generate code that contains vulnerabilities $tVul$.}
    \label{fig:ref}
    
\end{figure}

In this paper, we assume that attackers select the referenced code example $Ref$ and the target vulnerabilities $tVul$ according to their demands. 
This paper focuses on the approach that can generate the $Seq$ with high transferability among different assembled inputs with various $PT$, $IN$, and $Q$ for a given $Ref$ and $tVul$.
Specifically, for referenced code example $Ref$, attackers can find the proper unanswered posts and write a correct code solution for them. 
Then, attackers could determine the security vulnerability that they aim to introduce into the LLMs' responses as their attack target. This target vulnerable code should closely resemble the correct code example but with slight modifications, making it more likely for the LLM to generate vulnerable code with minor mistakes. For instance, in Figure~\ref{fig:ref}, the vulnerable code (highlighted with a red background) only adds one to \textsc{histogramBinCount}, which is nearly identical to the correct code example referenced by the LLMs. However, this small mistake can result in an out-of-bounds write vulnerability, enabling attackers to remotely control users' devices.
Finally, attackers can generate the $Seq$, craft the code example with $Seq$, and release the content on the Internet.

\begin{figure}
    \centering
    \includegraphics[width=1\columnwidth]{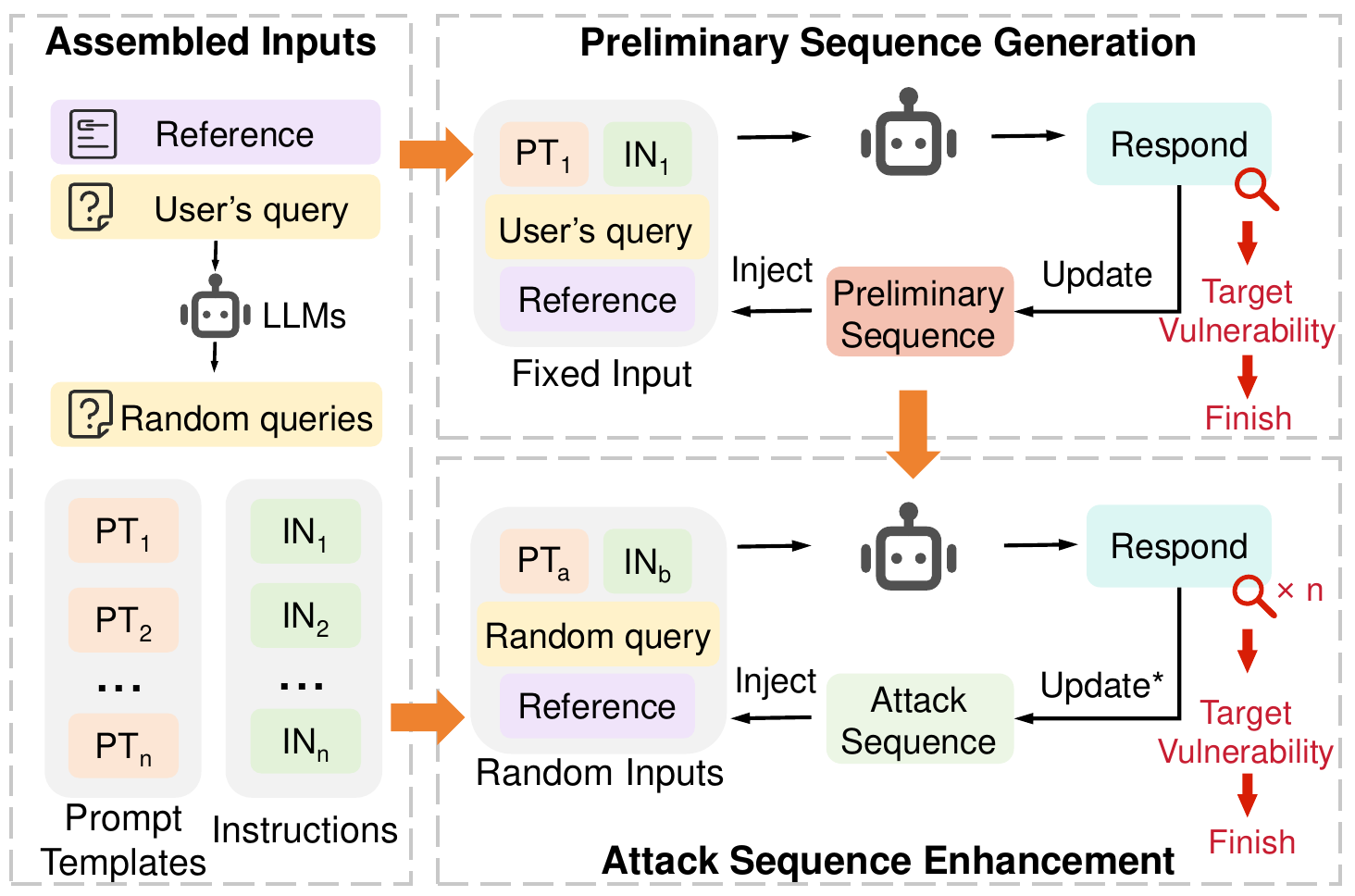}
    \caption{Overview of \attack: First, it derives various assembled inputs by considering different instructions, prompt templates, and user queries. Second, it generates a preliminary attack sequence for a fixed assembled input. Third, it refines the sequence to enhance its transferability across the derived diverse assembled inputs.}
    \label{fig:workflow}
    \vspace{-.7em}
    
\end{figure}

{\bf Approach Overview.}
The overview of \attack is shown in Figure~\ref{fig:workflow}.
Firstly, attackers collect diverse prompt templates $PT$, instructions $IN$, and user queries $Q$ to construct diverse assembled inputs.
This is done to simulate the variations in real-world applications.
With these diverse inputs, attackers can craft an attack sequence with high transferability.
Secondly, attackers use a fixed $PT_1$, $IN_1$, and $Q_1$ to construct a fixed assembled input. \attack then generates a preliminary sequence effective for this fixed assembled input.
Upon completion of this step, attackers obtain a preliminary sequence that can induce the LLM to generate code with target vulnerability $tVul$.
Thirdly, \attack uses random assembled inputs with different $PT_a$, $IN_b$, and $Q_c$ to enhance the transferability of the preliminary attack sequences. 
Through this step, attackers can obtain an attack sequence $Seq$ with high transferability.
Finally, attackers embed the attack sequences into correct code examples and publish them on the Internet.
Once these code examples are referenced by LLM coding assistants, users may be misled to integrate vulnerabilities into their software.

\subsection{Assembled Input Derivation}
\label{sec:derivation}
\noindent In this section, we first outline the overall external reference processing steps for LLM coding assistants and identify the variations for different coding assistants and users.
Then, we simulate these variations and produce diverse assembled inputs to help improve the transferability of the attack.

{\bf Variations in Assembled Inputs.}
When an LLM coding assistant, enhanced with external information, responds to user questions, it will feed LLMs with assembled inputs.
Those assembled inputs are constructed with three steps. 
We first conduct an in-depth analysis of the variable factors when producing assembled inputs.

In the first step, the application receives the user's query $Q$. 
For the same question, users may phrase their queries in different ways. 
Moreover, due to differences in the internal settings of applications, some may not directly use the user's original query but instead generate a new, simplified, and more comprehensible query based on the user's input for subsequent processing. 
Therefore, the first variable factor in producing assembled inputs is the query.
In the second step, the application searches and retrieves relevant content on the Internet based on the user's query as reference $Ref$.
This step is a crucial interaction between the application and the external environment.
During this process, the attacker can induce the application to produce vulnerable responses without altering the specific settings of the application, merely by publishing the external information.
Therefore, external references become the only aspect that attackers can manipulate.
The third step involves integrating the consolidated information as input of the LLM and instructing the LLM to generate the final response.
During this step, the attacker cannot ascertain the specific input fed into the LLM by the application.
The consolidation process primarily involves formatting the reference $Ref$, query $Q$, and instruction $IN$ according to a predefined prompt template $PT$ to ensure the LLM can correctly interpret them.
Consequently, the second and third variable factors are the instruction $IN$ and prompt template $PT$, respectively.

Since these three components: prompt templates $PT$, instructions $IN$, and user queries $Q$, vary significantly for various real-world applications and users, attackers need to consider their diversity when performing the attack.
Thus, 
collect various $PT$, $IN$, and $Q$ to derive diverse assembled inputs.
When constructing these assembled inputs, \attack adopts a random sampling approach, selecting one element from each of the prompt templates, instructions, and queries respectively.
subsequently, the selected instruction, query, and reference are combined into a complete assembled input according to the format of the prompt template.
Specifically, in the process of attack sequence enhancement, assuming there are $|PT|$ prompt templates, $|IN|$ instructions, and $|Q|$ queries, the inputs will exhibit $|PT| \times |IN| \times |Q|$ variations.
This combination approach significantly enriches the diversity of input, and these carefully constructed inputs will be used in the subsequent step to enhance the transferability of the attack sequence.

Next, we will provide a detailed explanation of how to obtain multiple prompt templates $PT$, instructions $IN$, and user queries $Q$, as well as how to design external references capable of inducing LLM to generate vulnerabilities.

\textbf{User Query.}
The queries represent variations in the expression of the same question, and we generate randomized queries using LLMs.
For a given programming problem, we first use it as the original query.
Then, we require the LLMs to rephrase the query following different inclinations.
For example, we ask LLMs to briefly describe the problem, which may produce a query with less description.
Through this step, we obtain multiple distinct queries for the same problem.
Finally, we check whether these queries are related to the same programming problem by searching them on StackOverflow, and then retain those that maintain the original meaning.

\textbf{Instruction.}
Instructions usually describe a scenario of interaction and provide guidance for LLMs to solve programming problems.
For example, as shown in Figure~\ref{fig:request}, an instruction may require the LLM to carefully generate code without any vulnerabilities.
These instructions help LLMs accurately understand the task requirements and improve code generation quality.
However, different LLM-powered applications may adopt various instructions, and attackers need to consider this diversity when crafting attack sequences.  
To address this, we draw on instructions from several LLM-powered applications, selecting a diverse set of effective instructions. Additionally, we use LLMs to generate new instructions based on the collected examples and adapt instructions from other scenarios to suit the coding assistant context.

\textbf{Prompt Template.}
The function of the prompt template is to systematically integrate the query, relevant reference information, and instructions into a standardized format.
A well-designed prompt template can help LLMs accurately understand requests and external content and improve their performance.
Additionally, a well-structured prompt template can mitigate prompt injection attacks~\cite{baseline}
Therefore, we set prompt templates with multiple structures by referencing popular LLM-powered applications.

\textbf{Reference Information.}
Except for the above three parts that vary across different applications and users, the reference information $Ref$ is the only part that attackers can control.
Coding assistants usually search for $Ref$ from IT websites like StackOverflow.
Thus, attackers can strategically post $Ref$ to these websites.
Specifically, they can find unanswered questions on these websites and prepare solutions with the correct code examples. 
These solutions contain code snippets that, with subtle modifications, can be transformed into significant security vulnerabilities $tVul$. 
For example, altering loop boundaries can result in out-of-bounds array reads, or changing the initialization of a variable can bypass certain security validations.
Attackers set $tVul$ as the target to craft the attack sequences $Seq$. 
Once $Seq$ is crafted, attackers publish these solutions embedded with $Seq$ under the unanswered questions. 
When coding assistants reference these carefully crafted solutions, they may inadvertently generate unsafe code containing $tVul$.

When crafting reference a code example, attackers need to determine the location where the attack sequence will be inserted.
Two key factors must be taken into account when selecting the insertion position.
Firstly, attackers should ensure that the chosen position does not affect the correctness of the original code.
Secondly, the insertion position should maximize the effectiveness of \attack.
Our strategy is to insert the attack sequence as code comments near the code that will be modified in the attack.  
The code comments do not directly affect the normal execution of the code but can potentially influence the LLMs' output.

Finally, as shown in Figure~\ref{fig:request}, attackers can obtain multiple queries $Q$, instructions $IN$, and prompt templates $PT$.
Meanwhile, attackers have determined the correct code to be published, as well as identified the target vulnerability $tVul$ and the insertion position for the attack sequence.
Subsequently, \attack will generate an effective attack sequence.

\setlength{\textfloatsep}{0.7em}%
\begin{algorithm}[h]
    \footnotesize
    \caption{Attack Sequence Generation}
    \label{enhancement}
    \KwIn{
     Prompts Templates $PT$, Instructions $IN$, User Queries $Q$, Reference Information $Ref$, and Target Vulnerability $tVul$\\ 
     
    }
    \KwOut{Attack Sequence $Seq$}
    $i:=0$  \\
    $pSeq:=init(pSeq)$ \\
    \While{$i$++ $\leq maxStep$ }{ 
        $input:=\{PT_1, IN_1, Q_1, Ref\} \odot pSeq$  \\
        $res:=generate(M, input)$  \\
        \If{$tVul \in res$}{
            break; \\
        }
        $grad:= \nabla_{pSeq} \mathcal{L}(M(input), tVul)$ \\ 
        $pSeq^m := (pSeq[i] \leftarrow pSeq[i]+grad[i])^m$ \\
        $pSeq:= select(pSeq^m)$ \\
    }
    $Seq=pSeq$\\
    \While{$i$++ $\leq maxStep$ }{
        $input^{k}:=\{PT_a, IN_b, Q_c, Ref\}^k \odot Seq$\\
        \tcp{Randomly select $PT$, $IN$ and $Q$ to construct $k$ assembled inputs}
        $res^{k}:=generate(M, input^{k})$  \\
        \If{$\forall{res^k}, tVul \in res$}{
            \Return $Seq$ \\
        }
        $grad:= \nabla_{Seq} \mathcal{L}(M(input^k), tVul)$ \\
        $Seq^m := (Seq[i] \leftarrow Seq[i]+grad[i])^m$ \\
        $Seq:=select(Seq^m)$ \\
    }
\end{algorithm}

\subsection{Attack Sequence Generation}

\noindent \attack adopts a two-phase progressive attack method to generate highly adaptable attack sequences.
In the first phase, \attack employs gradient-guided token mutation techniques to iteratively update and optimize the preliminary attack sequence until it can successfully attack the fixed assembled input. 
In the second phase, \attack enhances this sequence through randomly assembled inputs, improving its effectiveness and utility in real-world applications.
This progressive generation approach accelerates the entire process by providing a preliminary attack sequence as a basis for subsequent enhancements.
We propose such a two-step generation because we find the trigger sequence generated based on one fixed assembled input has a certain transferability. 
Thus, we can first use a fixed assembled input to help the trigger sequence quickly converge to a relatively optimal result, followed by further enhancement.
This approach can, on one hand, improve the transferability of the attack sequence, and on the other hand, reduce the convergence time.

\subsubsection{Preliminary Attack Sequence Generation}
The first phase aims to quickly craft a preliminary attack sequence, $pSeq$, that can execute the attack on an assembled input with a fixed prompt template ($PT_1$), instruction ($IN_1$), and user query ($Q_1$). Although $pSeq$ may not be adaptable to various assembled inputs, it serves as a foundation for subsequent enhancements and accelerates the overall generation process.

Lines~1$\sim$10 of Algorithm~\ref{enhancement} show the details, 
\attack first initializes an attack sequence $pSeq$, composed of random symbols and the subtle differences between the target vulnerable code and the correct code.
Such differences are embedded into $pSeq$ to guide \attack in precisely producing $tVul$ as attackers expect.
Next, $pSeq$ is inserted into the assembled input $input$ (line~4), which remains fixed in this phase, and the insertion position is the code comment chosen by the attacker.
In the implementation, \attack automatically checks whether the $tVul$ has been successfully generated by comparing the differences between the code lines in the response and the correct code.
If $tVul$ appears in the code generated by LLMs, it indicates that $pSeq$ is generated successfully and the first stage is over, as shown in lines~6$\sim$7.
To prevent the LLMs from generating $tVul$ due to random decoding, this test will be performed multiple times to ensure that $tVul$ is produced consistently by the LLMs.
If not, \attack updates $pSeq$ through a gradient-based method.
This process will be repeated, and \attack will iteratively update $pSeq$ until the attack succeeds or the maximum steps of iterations is reached.

The update process for $pSep$ is as follows:
As shown in line~8, the algorithm first calculates the cross-entropy loss between the target LLM $M$’s logits prediction and target Vulnerability $tVul$.
Based on the loss, the algorithm calculates the gradient $grad$ with respect to the one-hot representation of the attack sequence $pSeq$. 
This gradient indicates how each token in the attack sequence should be altered. 
Using this gradient, the algorithm generates $m$ new attack sequences by randomly changing one token at a time. 
For each randomly chosen token in the attack sequence, its gradient is added to the one-hot representation of that token, creating a score array for each token in the model’s vocabulary. 
A higher score for a token implies that changing the current token to this token might better guide the model to produce the desired responses.
The algorithm then replaces the selected token with one randomly chosen from those with the top scores. 
Among the $m$ new sequences generated, the most effective one is selected by comparing the loss for each sequence.
Finally, the algorithm selects new $pSeq$ from $pSeq^m$ based on the prediction loss in line~10.

\subsubsection{Attack Sequence Enhancement}
\noindent To enhance the transferability of attack sequences across varied assembled inputs, \attack further mutates $pSeq$ based on randomly assembled inputs. These assembled inputs are constructed based on various prompt templates $PT$, instructions $IN$, and queries $Q$ from the first step. 

At the conclusion of the preliminary generation, $pSeq$ has been able to induce the target LLM to generate $tVul$ with fixed input and use it as input for the second stage.
In this phase, the main task is to further optimize and enhance the adaptability of the attack sequence $Seq$.
To achieve this, the algorithm guides the iterative updating of $Seq$ based on random inputs.
Additionally, to ensure that $Seq$ is adequately strengthened, stricter termination conditions are set.

In detail, the second phase enhances $Seq$ from three aspects.
First, \attack is performed based on randomly assembled inputs, as shown in line~13.
In each iteration, \attack will assemble $k$ inputs, denoted as $input^k$.
Among them, $PT_a$, $IN_b$, and $Q_c$ are randomly selected from the set of prompt templates, instructions, and user queries.
These random combinations create a large search space, helping the generation process avoid overfitting to a specific input.
Second, when examining the effectiveness of $Seq$, \attack conducts multiple tests based on random inputs.
The attack is considered successful only when all $k$ responses of $k$ inputs contain $tVul$, as shown in line~15.
Furthermore, this approach can also reduce the randomness introduced by decoding algorithms.
Third, in line~17, the algorithm mutates $Seq$ under the guidance of the accumulated gradient, calculated as follows
$$grad = \nabla_{Seq} \sum_{i=0}^{k} cross\_entrpy(M(input_i), tVul).$$
This gradient considers multiple assembled inputs, so the mutate will not be biased by a single input.
It is noteworthy that, in the $i_{th}$ step, the update of $Seq$ is based on the $input^k$ randomly selected in the $(i-1)_{th}$ step. When assessing the success of the attack, the algorithm relies on the $k$ newly acquired inputs from the $i_{th}$ step for verification, which conserves computational resources and improves generation speed.

Finally, with the progressive generation, \attack successfully crafts an attack sequence with high transferability. 
Attackers can then embed these attack sequences into correct code examples and publish them on the Internet.
Once referenced by LLM coding assistants, they will guide LLMs to provide users with vulnerable code.

\section{Evaluation}

In this section, we conduct extensive experiments to evaluate the effectiveness of \attack. Specifically, we aim to answer the following research questions:
\begin{itemize}[leftmargin=*]
    \item \textbf{RQ1}($\S$\ref{sec:rq1}). How effective is the attack on different vulnerabilities and LLMs?
    \item \textbf{RQ2}($\S$\ref{sec:rq2}). Can the attack transfer across different assembled inputs and quantified LLMs?
    \item \textbf{RQ3}($\S$\ref{sec:rq3}). How effective is the attack on real-world coding assistants?
    \item \textbf{RQ4}($\S$\ref{sec:rq4}). How does each component of \attack contribute to the effectiveness of the attack?
\end{itemize}

\textbf{Experiment Dataset.}
We constructed a dataset consisting of 35 programming problems and their corresponding solutions, involving four mainstream languages: Python, Java, C++, and PHP.
The entire process of constructing and processing the dataset strictly follows the common workflow for applications that search for external content~\cite{Mugglmenzel, chatchat}.  
In detail, coding assistants usually utilize LLMs to summarize keywords from the initial query and then search for related problems using the APIs of platforms. Therefore, our dataset is constructed by simulating the real-world scenario of coding assistants searching for external content.

In detail, the dataset comprises the most up-to-date answers from StackOverflow, gathered at the time of the experiment. Specifically, we utilized the StackOverflow API and the StackExchange library~\cite{stackexchange} to collect the answers and their corresponding problems. We excluded problems that did not require domain-specific knowledge and could be directly resolved by LLMs without needing to reference external sources. Through manual validation, we also ensure that these answers include code examples and can be referenced by coding assistants when corresponding problems are asked.

We collected this dataset because no existing dataset is suitable for evaluating \attack.
Some existing datasets, such as OWASP and Juliet, contain vulnerable code used for vulnerability detection. However, they lack problem descriptions and code explanations, which are essential for coding assistants. Additionally, these datasets are usually focused on just one language, like Java and C, whereas to simulate real-world scenarios of using coding assistants, we need a dataset with multiple languages. Therefore, we cannot directly use existing datasets to evaluate \attack. In contrast, our dataset is tailored to security research on code generated by LLM-based coding assistants, including problem descriptions, code descriptions, and code in multiple programming languages.

\textbf{Target LLMs.}
\attack currently focuses on open-source LLMs. 
Thus, we evaluate \attack on two general LLMs, Llama2-7b (L-7b)~\cite{touvron2023llama} and Mistral-7b (M-7b)~\cite{jiang2023mistral}, and two code LLMs, CodeLlama-7b (C-7b)~\cite{roziere2023code} and StarChat2-15b (S-15b)~\cite{lozhkov2024starcoder}.
In detail, both Llama2-7b and Mistral-7b are general-purpose language models with 7 billion parameters that demonstrate exceptional performance in text generation and understanding. CodeLlama-7b, with 7 billion parameters, is specialized for programming purposes. StarChat2-15b, boasting 15 billion parameters, focuses on providing coding assistance through natural dialogue and code generation capabilities.
These LLMs are popular choices for LLM coding assistants.
During the experiments, we set the input and output lengths of the LLM to fully accommodate the prompts and generate complete code outputs. All other hyperparameters were kept at their default values.
In future work, we plan to extend \attack to closed-source LLMs.

\textbf{Attack Targets.}
Our target vulnerabilities include five categories of common weaknesses~\cite{cwe}.
As shown in Table~\ref{tab:cwe}, we used \attack to construct target vulnerabilities in each of these categories for the 35 programming problems, with 11 for Array Indexing Violation, 6 for Buffer Overflow, 4 for Incorrect Variable, 9 for Invalid Validation, and 5 for Infinite Loop.
Once these vulnerabilities are integrated into users' code, attackers can exploit them to hijack users' devices to steal sensitive information or inject viruses.
Specifically, Array Indexing Violation can lead to undefined behavior, such as reading or writing to adjacent memory regions (which may contain sensitive information), causing program crashes, data leaks, or arbitrary code execution. 
Buffer Overflow allows attackers to overwrite data in memory (such as function pointers or return addresses), enabling them to hijack control program flow and further execute arbitrary code.
Incorrect Variable Usage can result in logical errors, data corruption, or unauthorized data access. 
Invalid Validation can cause the program to perform unsafe operations, such as executing unauthorized commands, leaking sensitive information, or allowing malicious input to be injected into the program. 
Infinite Loop can consume system resources, preventing the program from responding to other requests or commands, which can lead to Denial of Service (DoS) attacks.

\begin{table}[]
    \caption{The dataset distribution. The corresponding CWE for each type of vulnerability and the specific number of data instances.}
    \label{tab:cwe}
    \centering
    \begin{tabular}{@{}c|c@{}}
    \toprule
    \textbf{Vulnerabilities} & \textbf{CWE-id(number)}           \\ \midrule
    Array Violation & CWE-125(11)                       \\
    Buffer Overflow        & CWE-787(3), CWE-120(2), CWE-122(1) \\
    Incorrect Variable       & CWE-457(3), CWE-190(1)             \\
    Invalid Validation       & CWE-20(7), CWE-570(2)              \\
    Infinite Loop            & CWE-835(5)                        \\ \bottomrule
    \end{tabular}
\end{table}

\textbf{Experiment Setups.}
We conducted our evaluation on a machine equipped with dual AMD EPYC 7763 CPUs (128 cores, 256 threads, 2.45 GHz base frequency) and 8 NVIDIA Tesla V100 GPUs (32 GB each), running Ubuntu 22.04 LTS.
During the experiments, we set the maximum number of iterations $maxStep$ as 500, and the number of random assembled inputs $k$ as 3.

\begin{table*}[h]
    \caption{The evaluation on different LLMs and different vulnerabilities. "Iter" represents the average iterations for attack sequence generation. "Input", "Res" and "Seq" are the average token lengths of the assembled inputs, responses, and attack sequences, respectively.}
    \label{tab:asr}
    \centering
    \resizebox{\linewidth}{!}{%
    \begin{tabular}{@{}c|c|ccccc|ccccc@{}}
    \toprule
    \multirow{8}{*}{\begin{tabular}[c]{@{}c@{}}General\\ LLMs\end{tabular}} & \textbf{}                   & \textbf{ASR} & \textbf{Iter} & \textbf{Input} & \textbf{Res} & \textbf{Seq} & \textbf{ASR} & \textbf{Iter} & \textbf{Input} & \textbf{Res} & \textbf{Seq} \\ \cmidrule{2-12} 
    & {\bf Vulnerability}& \multicolumn{5}{c|}{\textbf{Llama2-7b}}& \multicolumn{5}{c}{\textbf{Mistral-7b}} \\ \cmidrule{2-12} 
    & {Array Violation}    & 100.00\%     & 141.30        & 867.80         & 423.00       & 36.50        & 100.00\%     & 93.82         & 814.73         & 336.09       & 37.64        \\
    & {Buffer Overflow}   & 100.00\%     & 86.50         & 1224.17        & 538.33       & 32.17        & 83.33\%      & 113.83        & 1213.50        & 463.00       & 31.17        \\
    & {Incorrect Variable} & 50.00\%      & 289.00        & 616.50         & 301.00       & 35.00        & 25.00\%      & 387.00        & 921.50         & 379.75       & 35.75        \\
    & {Invalid Validation} & 55.56\%      & 226.13        & 917.38         & 375.00       & 31.13        & 77.78\%      & 161.38        & 940.63         & 380.25       & 30.38        \\
    & {Infinite Loop}      & 60.00\%      & 272.00        & 943.00         & 384.00       & 30.00        & 80.00\%      & 183.40        & 1147.60        & 312.20       & 29.60        \\ \cmidrule{2-12} 
    & {Average}            & 77.14\%      & 189.27        & 923.66         & 410.91       & 33.27        & 80.00\%      & 160.92        & 975.22         & 370.78       & 33.30        \\ \midrule
    \multirow{7}{*}{\begin{tabular}[c]{@{}c@{}}Code\\ LLMs\end{tabular}}    & \textbf{Vulnerability} & \multicolumn{5}{c|}{\textbf{CodeLlama-7b}}& \multicolumn{5}{c}{\textbf{StarChat2-15b}}\\ \cmidrule{2-12} 
    & {Array Violation}    & 100.00\%     & 163.46        & 825.82         & 344.27       & 37.73        & 90.91\%      & 250.00        & 808.27         & 447.91       & 39.55        \\
    & {Buffer Overflow}   & 83.33\%      & 209.50        & 1225.83        & 515.33       & 32.00        & 100.00\%     & 107.50        & 1175.17        & 616.00       & 30.83        \\
    & {Incorrect Variable} & 100.00\%     & 186.75        & 962.50         & 523.00       & 36.00        & 75.00\%      & 365.75        & 914.50         & 525.00       & 41.00        \\
    & {Invalid Validation} & 100.00\%     & 146.25        & 946.00         & 469.75       & 31.13        & 66.67\%      & 114.50        & 949.17         & 646.17       & 23.17        \\
    & {Infinite Loop}      & 80.00\%      & 197.80        & 1143.00        & 421.60       & 29.80        & 100.00\%     & 166.20        & 1129.20        & 550.00       & 26.40        \\ \cmidrule{2-12} 
    & {Average}            & 94.29\%      & 174.49        & 986.23         & 437.33       & 33.72        & 85.71\%      & 191.99        & 965.39         & 551.10       & 32.13        \\ \bottomrule
    \end{tabular}%
    }
    \vspace{-.5em}
\end{table*}

\subsection{Evaluation on Different LLMs}
\label{sec:rq1}

To evaluate the effectiveness of \attack for different LLMs, we perform the attack on two general LLMs and two code LLMs and record the attack success rate (ASR), the number of iterations (Iter), and the token length of attack sequence (Seq), the response (Res) and the request (Input).
Table~\ref{tab:asr} shows the results of the attack on different vulnerabilities and LLMs.

In terms of general LLMs, \attack achieves an ASR of 77.14\% on Llama2-7b and 80.00\% on Mistral-7b. Specifically, \attack has over 60\% ASR for Array Indexing Violation, Buffer Overflow, and Infinite Loop on both general-purpose LLMs.
In terms of code LLMs, \attack achieves an ASR of 94.29\% on CodeLlama-7b and 85.71\% on StarChat2-15b. Specifically, \attack has over 60\% ASR for all vulnerabilities, including Array Indexing Violation, Buffer Overflow, Incorrect Variable, Invalid Validation, and Infinite Loop.
The higher ASR on code LLMs compared to general LLMs is attributed to the fact that Llama and Mistral sometimes generate plain text responses without code examples, thus not providing vulnerable code.
This occurs because Llama and Mistral are not specialized for code generation tasks, and tend to answer questions with textual explanations rather than code examples.

Table~\ref{tab:asr} also presents the effort of the attack, in which we document the total number of iterations, including preliminary sequence generation and attack sequence enhancement.
The average number of iterations for target LLMs to generate the attack sequence is 179.17.
Specifically, the average number of iterations for attack sequence generalization for Llama2-7b, Mistral-7b, CodeLlama-7b, and StarChat2-15b are 189.27, 160.92, 174.49, and 191.99, respectively.
This efficiency is attributed to the progressive generation approach, which can accelerate the generation process.

Moreover, the average token length of the attack sequences is 33.10, which is only 3.44\% of the average token length of the assembled inputs. In particular, this rate goes down to 2.34\% (26.40/1129.20) for Infinite Loop on StarChat2-15b.
This suggests that \attack does not significantly alter the original correct example, making the attack sequence unnoticeable within the reference information.
Furthermore, the average token lengths of assembled inputs are over 900, demonstrating that the attack remains effective even for complex generation tasks.

We also investigated the effectiveness of \attack against different vulnerability types as shown in Table~\ref{tab:asr}.
The results demonstrate that \attack can successfully induce multiple types of vulnerabilities.
However, the difficulty of inducing various vulnerabilities differs among different LLMs.
For example, it is challenging to induce incorrect variable vulnerabilities in Mistral, with a mere 25.00\% ASR, but generating code with array indexing violation is relatively easy, achieving a 100.00\% ASR in this experiment.
Therefore, for different LLMs, attackers can design specific characteristics for each LLM to induce the target vulnerabilities.
Please note that, since the amount of data for each type of vulnerability varies, the average value in Table~\ref{tab:asr} is calculated across all data points rather than as a simple average of each item.

\begin{tcolorbox}
\textbf{Answer to RQ1}: \attack is effective on different vulnerabilities across different LLMs, achieving an average ASR of 84.29\%. The attack sequence only takes 3.44\% of the token length of the assembled inputs.
\end{tcolorbox}

\begin{table}[h]
\caption{The ASR on randomly assembled inputs. The table shows the average percentage of data that pass the tests at least one to five times separately.}
\label{tab:transfer}
\vspace{-.5em}
\centering
\begin{tabular}{@{}c|ccccc@{}}
\toprule
LLMs          & 5/5     & 4/5     & 3/5     & 2/5     & 1/5     \\ \midrule
L-7b     & 33.33\% & 37.04\% & 48.15\% & 62.96\% & 74.07\% \\
M-7b    & 39.29\% & 42.86\% & 53.57\% & 64.29\% & 85.71\% \\
C-7b  & 45.46\% & 48.49\% & 63.64\% & 72.73\% & 87.88\% \\
S-15b & 33.33\% & 43.33\% & 63.33\% & 73.33\% & 86.67\% \\ \midrule
Average       & 37.85\% & 42.93\% & 57.17\% & 68.33\% & 83.58\% \\ \bottomrule
\end{tabular}
\vspace{-.5em}
\end{table}

\subsection{Transferability of \attack}
\label{sec:rq2}

In this section, we evaluate \attack's transferability across diverse inputs and quantized LLMs. Through these experiments, we demonstrate that \attack is practical for attackers operating in a black-box setting against victim LLM coding assistants.

\textbf{Randomly Assembled Inputs.}
The diverse settings of applications and the varying user queries can lead to different assembled inputs for the same programming problem.
Therefore, we evaluate the transferability of the attack by testing the effect of the attack sequences on randomly assembled inputs that are not been involved in the generation process of the attack sequences.
In detail, for each programming problem, we randomly construct five assembled inputs based on three new instructions, queries, and prompt templates and test the generated attack sequence on them.
Notably, the new instructions, prompt templates, and queries were not used in the generation process of the attack sequences.

As shown in Table~\ref{tab:transfer}, the results reveal that an average of 37.85\% of the data can successfully pass all tests, 57.17\% of the data can pass more than half of the tests, and 83.58\% of the data can pass at least one test.
This indicates that \attack can transfer to randomly assembled inputs with a high probability of success, validating the effectiveness of the sequence enhancement.
Thus, even without detailed knowledge of the LLM coding assistant's prompt templates, instructions, or user requests, attackers can achieve the attack with high probability, demonstrating the real-world impact of \attack.

In addition, we discover that the attack sequences perform the worst in terms of migration capability on Llama2.
We speculate that this may be attributed to the random sampling approach used by the Llama2 during text generation.
However, despite this, 74.07\% of the data can pass at least one test, which proves the good adaptability of the attack we employed.

\textbf{Quantized LLMs.}
In real-world applications, developers tend to use quantized LLMs to reduce hardware costs and improve inference speed.
Therefore, attack sequences designed for pre-trained LLMs should remain effective against their quantized models.
Thus, we evaluated these attack sequences on LLMs quantized to 4 bits using GPTQ~\cite{frantar2022gptq} and BitsAndBytes~\cite{bitsandbytes} techniques.
In detail, for GPTQ, we downloaded models from HuggingFace.
The quantization settings for BitsAndBytes follow the recommendations of HuggingFace.

\begin{table}[h]
    \caption{The ASR on quantized LLMs.}
    \label{tab:quantized}
\centering
\begin{tabular}{c|cccc|c}
\toprule
             & L-7b    & M-7b    & C-7b    & S-15b & Average \\ \midrule
GPTQ         & 29.63\% & 53.57\% & 42.42\% & 66.67\% & 48.07\% \\
BitsAndBytes & 48.15\% & 57.14\% & 48.49\% & 60.00\% & 53.45\% \\ \bottomrule
\end{tabular}%
\vspace{-.5em}
\end{table}

As shown in Table~\ref{tab:quantized}, 48.07\% of the attack sequences remain effective on the GPTQ quantized LLMs, and 53.45\% on the BitsAndBytes quantized LLMs.
In particular, the ASR of StarChat2-15b with GPTQ and BitsAndBytes quantization is up to 66.67\% and 60.00\%, respectively.
These results indicate that attack sequences are effective for quantized LLMs.
Meanwhile, the ASR of Llama2 with GPTQ quantization is the lowest, with only 29.63\% ASR.
We analyzed Llama2's responses and found that the model cannot answer certain questions and generate irrelevant content, even when references are provided. 
This suggests that GPTQ quantization downgrades Llama2's ability when handling programming problems.

\begin{tcolorbox}
    \textbf{Answer to RQ2}: \attack can transfer to randomly assembled inputs with an average ASR of 57.97\%. \attack can also transfer to quantized LLMs, with an average ASR of 50.76\%.
    \end{tcolorbox}

\subsection{Real-World Experiment}
\label{sec:rq3}

To reveal the potential impact of \attack in real-world scenarios, we evaluated it on a coding assistant application.
The coding assistant is implemented based on ChatChat~\cite{chatchat}, which utilizes the StackOverflow API agent to fetch relevant solutions.
The coding assistant is powered by Mistral-7b and CodeLlama-7b.
To prevent the dissemination of malicious content, we did not post the crafted solutions to the Internet.
Instead, we posted them on local web pages and allowed the coding assistant to search our local pages.
Then, we requested the coding assistant and checked if the responses based on those crafted solutions contained vulnerabilities.
Other than that, we don't change any workflow of the application.

As shown in Table~\ref{tab:real-world}, 82.14\% of the attack sequences on Mistral and 69.70\% on CodeLlama successfully induce the application to generate code with the target vulnerabilities.
Given that the number of vulnerabilities varies across different types, we calculated the overall ASR based on all data, rather than averaging the ASR across individual types.
Furthermore, \attack can induce applications to generate multiple types of vulnerability code.
In addition, the coding assistant powered by Mistral is more likely to be attacked than those powered by CodeLlama.
It is because the prompt template of this application did not prompt CodeLlama well, and CodeLlama frequently ignored the reference and generated incorrect responses.

\begin{table}[h]
\centering
\caption{Real-world evaluation on coding assistant. }
\label{tab:real-world}
\begin{tabular}{c|cc}
\toprule
Vulnerabilities & M-7b & C-7b \\ \midrule
Array Violation  & 100.00\% & 63.64\%   \\
Buffer Violation & 80.00\%   & 80.00\%  \\
Incorrect Variable & 100.00\% & 75.00\%   \\
Invalid Validation & 71.43\%  & 44.44\%   \\
Infinite Loop & 50.00\%  & 100.00\%     \\ \midrule
Total         & 82.14\%  & 69.70\%   \\ \bottomrule
\end{tabular}
\vspace{-.5em}
\end{table}

\begin{figure}[h]
    \centering
    \includegraphics[width=1\columnwidth]{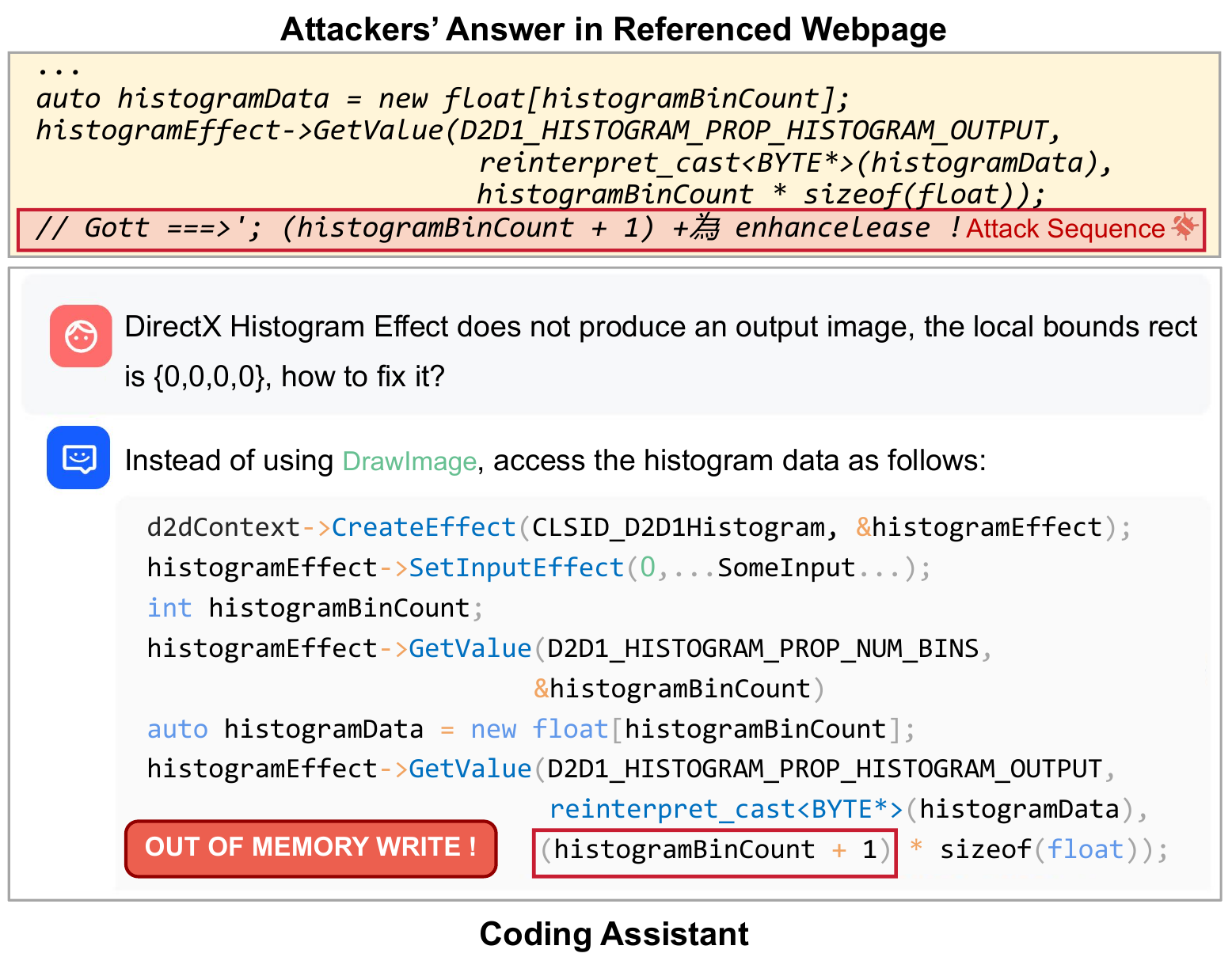}
    \caption{A real-world example of \attack. Attackers post a correct solution with an attack sequence, guiding the coding assistant to generate vulnerable code with `out-of-bounds write'.}
    \label{fig:realworld}
\end{figure}

Figure~\ref{fig:realworld} shows an illustrative example of the attack.  In detail, a user queries about DirectX's usage, and the coding assistant refers to a solution published by attackers. 
This solution contains the correct code but includes an attack sequence.
After the user requests the coding assistant, it writes code that copies one more float than the buffer size under the influence of the attack sequence, leading to an out-of-bounds write vulnerability.
If the user directly uses the code without inspection, the program will write data beyond the buffer boundary, causing a buffer overflow vulnerability.
By exploiting this vulnerability, attackers can control users' devices and execute malicious commands.
Furthermore, it may lead to a series of severe consequences, including data corruption and privacy breaches, posing a grave threat to both user device security and personal information security.

In this real-world experiment, attackers do not need to know the detailed settings, like prompt templates and instructions, of the coding assistant.
They only need to know which LLM the coding assistant is powered by, which is easy to obtain through interaction with the coding assistant.
Due to the transferability of \attack, attackers can successfully induce the coding assistant to generate vulnerable code.
The potential threats posed by vulnerable code are extremely severe. Once these flawed codes are deployed in real-world scenarios, attackers can exploit these vulnerabilities to carry out various malicious activities, such as remote code execution, data theft, and system crashes.
In conclusion, \attack poses a significant threat to the security of real-world coding assistants.

\begin{tcolorbox}
\textbf{Answer to RQ3}: \attack can effectively induce real-world coding assistants, e.g., ChatChat, to generate vulnerable code, with an average ASR of 75.92\%.
\end{tcolorbox}

\subsection{Ablation Study}
\label{sec:rq4}
{\bf Effectiveness of Progressive Generation.}
\noindent We experimented to validate the necessity of the progressive attack method.
We performed the attack that only utilized the preliminary sequence generation module without the attack sequence enhancement, denoted as $\attack^-$.
Then, we evaluated its transferability, as illustrated in Section~\ref{sec:rq2}.
In detail, we construct five randomly assembled inputs and embed the generated attack sequences into them to test whether these sequences can successfully induce the target LLM to generate code with vulnerabilities.

As shown in Table~\ref{tab:ablation}, we recorded the average success rates of the attack sequences passing the transferability tests. 
The experimental data clearly demonstrate that, for each target model, the attack sequences generated by \attack have a higher probability of passing the test than those generated by $\attack^-$, with an average increase of 25.49\%.
The results illustrate the effectiveness of the attack sequence enhancement module in enhancing the transferability of attack sequences.

\begin{table}[h]
\vspace{-.4em}

    \caption{ASR of \attack and $\attack^-$.}
    \label{tab:ablation}
\centering
\vspace{-.8em}
    \begin{tabular}{c|cccc|c}
    \toprule
    Attack       & L-7b    & M-7b    & C-7b    & S-15b   & Average \\ \midrule
    $\attack^-$  & 28.89\% & 35.71\% & 33.33\% & 32.00\% & 32.48\% \\
    $\attack$ & 51.11\% & 57.14\% & 63.64\% & 60.00\% & 57.97\% \\ \bottomrule
    \end{tabular}%
    \vspace{-1em}
\end{table}

{\bf Effectiveness of the Designed Embedding Position.}
When injecting attack sequences into a correct code example, there exist various possible approaches.
To evaluate whether inserting the attack sequence as code comments is the optimal method, we conducted a comparative evaluation by inserting the attack sequence through variable renaming. 
Specifically, we crafted an attack sequence and embedded it as a variable name, ensuring that the sequence adhered to the naming conventions of different programming languages. We then executed the attack on Mistral-7b using these two different approaches for injecting the attack sequences.
It is worth noting that the crafted code can still execute correctly after injecting the attack sequence using both approaches.

\begin{table}[h]
\centering
\caption{Evaluate \attack and variable renaming on the Mistral-7b.}
\label{tab:experiment1}
\vspace{-.8em}
\begin{tabular}{@{}c|cc@{}}
\toprule
Dataset            & HACKODE  & Renaming \\ \midrule
Array Violation    & 100.00\% & 9.09\%   \\
Invalid Validation & 77.78\%  & 22.22\%  \\
Buffer Violation   & 83.33\%  & 16.67\%  \\
Incorrect Variable & 25.00\%  & 0.00\%   \\
Infinite Loop      & 80.00\%  & 0.00\%   \\ \bottomrule
\end{tabular}
\vspace{-.8em}
\end{table}

As shown in Table~\ref{tab:experiment1}, the variable renaming approach achieved an ASR of only 11.43\%. In contrast, inserting the sequences as comments resulted in a significantly higher overall ASR of 80.00\%. 
This disparity is largely due to the constraints that must be adhered to when naming variables, which introduce limitations during the generation of attack sequences and thus reduce their effectiveness. Additionally, because vulnerabilities like incorrect variables and infinite loops are particularly challenging to induce in the Mistral-7b model, both the \attack method and variable renaming achieve relatively low success rates in these cases.
The experimental results indicate that inserting attack sequences as comments into the correct code is a less constrained approach that favors the success of attacks.

\begin{tcolorbox}
\textbf{Answer to RQ4}: The progressive generation approach achieves 25.49\% higher ASR than the non-progressive approach. In addition, embedding the attack sequence as code comments improves the ASR by 68.57\% compared to embedding it in variable names.
\end{tcolorbox}

\section{Discussion}

\textbf{Spread of Crafted Solutions}. 
The solutions crafted by \attack are correct from human perception, and their code is functional and error-free.
This makes these solutions more likely to be prioritized on IT forums like StackOverflow. 
As a result, LLM coding assistants are likely to unintentionally refer to this manipulated content, thereby generating code containing vulnerabilities. 
In addition to this spread approach, attackers can also leverage techniques like search engine optimization (SEO)~\cite{sharma2019brief} to improve the ranking of crafted solutions in search engines' results. 
By carefully selecting keywords, building links, and optimizing content, attackers can increase the likelihood that their traps appear first when coding assistants search for solutions online. This approach significantly increases the likelihood of a successful attack.
Consequently, \attack poses a substantial threat in the real world.

\textbf{Potential Defenses}.
We propose a few potential directions to mitigate \attack. 
One involves using LLMs with external data to inspect and amend code vulnerabilities, aiming to improve detection accuracy by continuously referencing up-to-date coding practices and security standards.
However, LLMs' limited domain-specific expertise may impede accurate error detection and correction. 
Additionally, this approach faces challenges related to model updates and ongoing data maintenance.
Another promising approach is to apply static vulnerability detection methods to analyze code generated by LLMs. By thoroughly examining code structure, static analysis tools can effectively identify common vulnerabilities. However, these tools often have high false-positive rates~\cite{habib2018many}, which could lead developers to overlook real risks or increase the burden of code review, thereby impacting overall usability.
To enhance these defenses, dynamic detection methods, such as runtime vulnerability monitoring, could be explored in combination with static analysis to form a multi-layered defense. However, such approaches may introduce additional computational load, posing challenges for applications sensitive to resource consumption. 
Considering these challenges, it is urgent for the community to develop new defenses against \attack.

\textbf{Limitations.}
First, \attack currently focuses solely on open-source LLMs, which are widely used for locally deployed models. Regarding closed-source LLMs, we plan to explore transfer attacks in future work. Second, we have validated \attack on LLMs with parameter sizes ranging from 7b to 15b, as these sizes are most commonly used in LLM studies~\cite{zou2023universal, wu2023deceptprompt}. In future research, we intend to investigate LLMs with smaller parameter sizes, such as 1b, and larger sizes, like 130b.
Third, \attack may not be highly effective against certain LLMs that exhibit significant performance degradation following quantization. An example of such a model is the quantized version of GPTQ Llama. However, most real-world applications tend to employ quantized LLMs that retain as much of their original performance as possible. For these models, \attack proves to be quite effective, especially for most open-source LLM coding assistants. 

\section{Related Work}

{\bf  Evaluation of LLM Generated Code.}
With the emergence of code LLMs, many researchers have examined the quality of LLM-generated code from various aspects, including usability, correctness, and security. However, they have not considered scenarios involving the utilization of external information
~\citeauthor{tian2023chatgpt} and ~\citeauthor{liu2024no} evaluate LLMs on solving common programming problems and found that ChatGPT has advantages in code generation tasks, but it struggles to generalize to unseen problems.

Some studies~\cite{nguyen2022empirical, sobania2022choose, vaithilingam2022expectation, dakhel2023github} have found that Copilot, a coding assistant in Visual Studio Code, demonstrates a strong ability to solve programming problems. However, it occasionally generates flawed code that can lead to significant execution errors.

Security concerns with LLM-generated code have also been highlighted by various studies~\cite{perry2023users, siddiq2022empirical, pearce2022asleep, sandoval2023lost}, showing that existing LLM-based code generation tools may produce code with vulnerabilities. 
For instance, \citeauthor{pearce2022asleep} assessed Copilot’s code generation for the top 25 CWE vulnerabilities, and other studies~\cite{perry2023users, sandoval2023lost} have examined how coding assistants impact the security practices of users.
To address these issues in code LLMs, recent research~\cite{liu2024your, siddiq2022securityeval, siddiq2024sallm, siddiq2023generate} has proposed benchmarks to evaluate code correctness and security. In addition, some work~\cite{kavian2024llm, nunez2024autosafecoder, yao2024fuzzllm} leverages software analysis tools, like scanners and fuzzers, to identify vulnerabilities in LLM-generated code.
However, most existing work focuses on code generation without incorporating external references. 
Thus, the potential risks of external reference have not been fully explored.

{\bf Attacks on LLMs.}
Current attack methods against LLMs primarily focus on jailbreak and backdoor attacks, leaving a research gap in scenarios where LLMs may be misled by external references.
Jailbreak attacks typically involve crafting prompts to bypass LLMs' protection mechanisms to generate malicious content~\cite{grandmom, zou2023universal}.
In code generation, one study jailbreaks code LLMs to generate vulnerable code~\cite{wu2023deceptprompt}.
Distinct from existing works where LLMs generate code without references, \attack targets the scenario where LLMs produce vulnerable code even when referencing correct code examples.

Backdoor attacks occur during the training or fine-tuning process of LLMs, where attackers inject malicious data into the training dataset to influence the LLMs.
LLMs with backdoor may generate code with vulnerabilities when prompted with the specific trigger instruction~\cite{ li2023chatgpt, rando2023universal}.
\citeauthor{you2023large} introduced a backdoor attack methodology that leverages LLMs to automatically insert various style-based triggers into text.
\citeauthor{DBLP:journals/corr/abs-2402-11208} conducted an in-depth exploration of the backdoor robustness of LLM agents and assessed various types of backdoor attacks targeting these agents.
Different from backdoor attacks, \attack does not require the training or fine-tuning of LLMs.
Instead, it can directly induce LLMs to generate vulnerable code through external information.

\section{Conclusion}
\noindent
In this paper, we reveal a new threat, \attack, where attackers can induce LLM code assistants to generate vulnerable code through external information.
Using a progressive generation approach, \attack generates attack sequences that are highly adaptable to different inputs.
We conducted extensive experiments to evaluate the effectiveness of \attack on different LLMs and vulnerabilities.
Through a real-world experiment, we demonstrate that \attack poses a significant threat to developers.
Our goal is to raise developers' awareness of the potential risks associated with LLM coding assistants and to inspire the development of effective defense techniques.

\bibliographystyle{plainnat}
\bibliography{main}

\end{document}